# On the Beam Profile and Beam Quality of Gain-Guided Index-Antiguided Fibers with Finite Cladding Boundary

Parisa Gandomkar Yarandi, Krishna Mohan Gundu, and Arash Mafi [1]

Department of Electrical Engineering and Computer Science
University of Wisconsin-Milwaukee
Milwaukee, WI, 53211

**Abstract:** The beam quality factor $M^2$ for the fundamental LP01 mode of a step-index fiber with a finite cladding diameter is calculated in the presence of gain, in a closed form, as a function of the complex generalized V-number. It is shown that the conventional infinite cladding diameter approximation cannot be used for index-antiguided gain-guided fibers, and the rigorous analysis presented in this paper is required for accurate prediction of the beam quality factor, as reported in recent experimental measurements.

---

[1] e-mail: `mafi@uwm.edu`

1. **Introduction**

It is highly desirable to increase the core size of single-mode fiber lasers and amplifiers. A large core diameter can help mitigate the unwanted nonlinear optical effects, raise the optical damage threshold, and increase the amplification per unit length of the fiber [1]. Siegman [2,3] has recently shown that gain-guided, index-antiguided (GG-IAG) fibers can operate in a robust single transverse mode, even for arbitrarily large core diameters. Since then, several experiments have demonstrated GG+IAG in various fiber laser configurations [4–8]. Single-mode operation in optical fibers with unprecedented core diameters of larger than 200 $\mu m$ has been observed.

In his landmark papers, Siegman carried out a comprehensive analysis of GG+IAG fibers, assuming an infinite diameter for the cladding [2, 3]. While the infinite-cladding is an acceptable approximation in conventional single-mode index-guiding (IG) fibers, we suggest that the finiteness of the cladding diameter can have a considerable impact of the physical characteristics of the propagating beam in GG+IAG fibers and related experimental measurements. There are two main reasons for the inadequacy of the infinite-cladding approximation. One is that due to practical limitations on the total diameter of the fiber and because of the extremely large core sizes, all existing experiments [4–8] report unusually small ratio of the cladding to core diameter. Another is that the GG+IAG mode, characteristically, can have a very long and slowly decaying intensity profile in the cladding, extending all the way to the outer boundary of the cladding. For example, we recently showed the inadequacy of the infinite-cladding approximation in predicting the beam quality measurements in various experiments on GG+IAG fibers [9].

Since the cladding to core diameter ratio is likely to remain small in practical GG+IAG designs, a detailed analysis of the impact of the cladding truncation on these fibers is warranted. In this paper, we build upon the results of Siegman in Refs. [2, 3] and extend his work to the case where the cladding of the GG+IAG fiber has a finite diameter and is truncated by a jacket. The jacket is assumed to be infinite and can be a polymer or glass protective layer on the cladding, or simply air. We explore the implications of the finite cladding diameter in detail and illustrate its main similarities and differences with the case of infinite cladding (IC) diameter. In particular, we show that a reliable comparison with the beam-quality $M^2$ measurements in recent experiments is only possible if the presence of the cladding-jacket interface is explicitly taken into account.

In subsection 2.2.A, we will present a brief overview of the GG+IAG fibers, for the ideal case where the cladding extends to infinity. In subsection 2.2.B, we extend the formalism to the case of finite cladding. The results will be used in sections 3 and 4 to study the impact of finite cladding on the beam profile and beam quality of GG+IAG optical fibers, respectively. We will conclude in section 5. Finally, in the Appendix, we will report on an analytical derivation of the $M^2$ in GG+IAG fibers. Proper care has been taken in choosing a notation that conforms with that of Siegman [3], in order to clearly illuminate the transition from the finite-cladding formulation to the limiting case of the infinite-cladding.

## 2. General Characteristics

### 2.A. GG+IAG fiber with infinite cladding

In the presence of gain, an ideal (GG+IG or GG+IAG) step-index optical fiber with infinite cladding can be characterized by a generalized complex V parameter squared [2] defined as

$$\tilde{V}_{12}^2 = \Delta N + iG. \tag{1}$$

The index and gain parameters $\Delta N$ and $G$ are given by

$$\Delta N = \left(\frac{2\pi a}{\lambda}\right)^2 2n_2 \Delta n, \tag{2}$$

$$G = \left(\frac{2\pi a}{\lambda}\right)^2 \left(\frac{n_2 \lambda}{2\pi}\right) g, \tag{3}$$

where $n_2$ is the refractive index of the cladding. $n_1 = n_2 + \Delta n$ is the refractive index of the core, $a$ is the core radius, $g$ is core power-gain coefficient, and $\lambda$ is the vacuum wavelength. $\Delta N$ is negative for an IAG fiber. For a proper choice of $\Delta N$ and $G$, Siegman [2] has shown that the core can support an LP01 guided mode in the form of

$$E(x, y, z_0) = \begin{cases} \tilde{N} \dfrac{J_0(u_1 r/a)}{J_0(u_1)} & r \leq a, \\ \tilde{N} \dfrac{K_0(u_2 r/a)}{K_0(u_2)} & a < r. \end{cases} \tag{4}$$

$\tilde{N}$ is an overall constant to be determined from the normalization condition (5) assumed throughout this paper,

$$\iint dS\, |E(x, y, z)|^2 = 1, \tag{5}$$

where $\iint dS \stackrel{\text{def}}{=} \iint dxdy$. The complex modal parameters $u_1$ and $u_2$ are defined as

$$u_1^2 = (n_1^2 k_0^2 - \beta^2)a^2, \tag{6}$$
$$u_2^2 = (\beta^2 - n_2^2 k_0^2)a^2, \tag{7}$$

where $\beta$ is the propagation constant of the propagating mode, and $k_0 = 2\pi/\lambda$. The modal parameters are related to the $V$ number as

$$u_1^2 + u_2^2 = \tilde{V}_{12}^2 = \Delta N + iG. \tag{8}$$

The dispersion Eq. 9 is obtained from matching the electric field and its slope at the core-cladding interface.

$$\frac{u_1 J_1(u_1)}{J_0(u_1)} = \frac{u_2 K_1(u_2)}{K_0(u_2)}. \tag{9}$$

Eqs. 8 and 9 can be used to determine $u_1$ and $u_2$, given $\tilde{V}_{12}$. In general, the value of $\tilde{V}_{12}$ determines the total number of confined guided modes in the fiber in the presence of gain, which can be zero or higher. We note that a confined propagation mode must decay exponentially in the radial direction in the cladding. This condition requires the complex modal parameter $u_2$ to have a positive real value ($\Re(u_2) > 0$). Those solutions of Eqs. 8 and 9 that satisfy $\Re(u_2) < 0$ are the leaky modes that grow exponentially to infinity in the transverse direction in the cladding. In the complex plane of ($\Delta N, G$) shown in Fig. 1, $\Re(u_2) = 0$ serves as the threshold line between confined and leaky modes and is portrayed as a solid line. The confined guiding modes are related to the points above the solid threshold line, while the leaky modes relate to the points below the threshold line. Here, and throughout this paper, we restrict our studies to the more interesting IAG region where $\Delta N < 0$, since the properties of the IG region where $\Delta N > 0$, are not much different from a conventional optical fiber in the absence of gain.

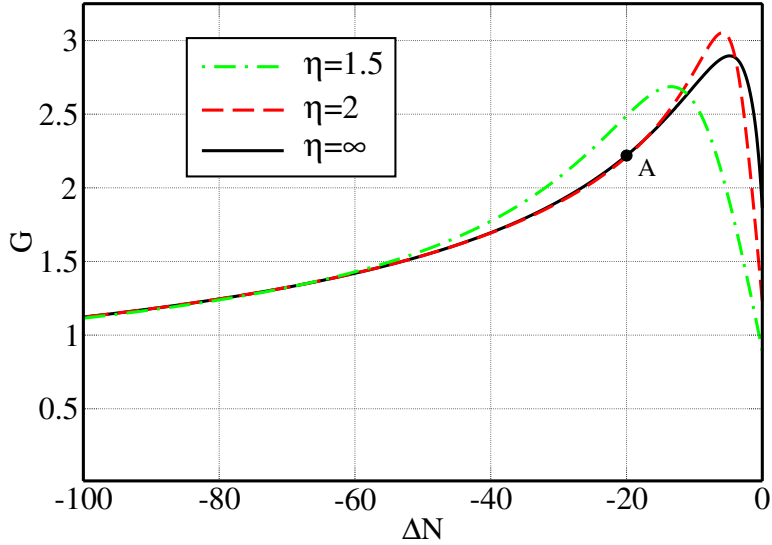

Fig. 1. The lines represent $\Re(u_2) = 0$ for the case of IC-GG fibers ($\eta = \infty$) and FC-GG fibers ($\eta = 2,\ 1.5$). For $\eta = \infty$, the solid line serves as the mode boundary threshold separating the LP01 guiding modes from the leaky modes in the complex ($\Delta N, G$) plane.

### 2.B. GG+IAG fiber with finite cladding

In this subsection, we generalize the formulation of GG optical fibers presented in subsection 2.2.A to the case where the cladding has a finite radius $b$. The refractive index profile of the finite-cladding GG fiber (FC-GG) is given by

$$n(r) = \begin{cases} n_1 & r \leq a, \\ n_2 & a < r \leq b, \\ n_3 & b < r. \end{cases} \quad (10)$$

$n_3$ is the refractive index of the medium surrounding the cladding, referred to as the jacket. The electric field profile of the LP01 core guided mode can be generally expressed as

$$E(x, y, z_0) = \begin{cases} \tilde{N} \dfrac{J_0(u_1 r/a)}{J_0(u_1)} & r \leq a, \\ \tilde{N} \dfrac{L_0(u_2 r/a)}{L_0(u_2)} & a < r \leq b, \\ \tilde{N} \dfrac{L_0(u_2 \eta)}{L_0(u_2)} \dfrac{K_0(u_3 r/b)}{K_0(u_3)} & b < r. \end{cases} \quad (11)$$

$\tilde{N}$ is an overall constant to be determined from the normalization condition (5). We have also defined

$$L_0(x) := K_0(x) + (S_2/S_1) J_0(x). \quad (12)$$

$\eta = b/a$ is the ratio of the radius of the cladding to the radius of the core, and $S_1$ and $S_2$ are defined in Eqs. 13 and 14.

$$S_1 = u_1 J_0(u_2) J_1(u_1) - u_2 J_0(u_1) J_1(u_2), \quad (13)$$
$$S_2 = u_2 J_0(u_1) K_1(u_2) - u_1 K_0(u_2) J_1(u_1). \quad (14)$$

$u_1$, $u_2$ and $u_3$ are the modal parameters in the core, cladding, and the outer layer, respectively. $u_1$ and $u_2$ are defined in Eqs. 6 and 7, while $u_3$ is defined in Eq. 15

$$u_3^2 = (\beta^2 - n_3^2 k_0^2) b^2. \quad (15)$$

The continuity of the field and its first derivative across the boundaries, $r = a$ and $r = b$, results in the following dispersion relation

$$\left( \eta \frac{u_2}{u_3} \right) \frac{S_1 K_1(u_2 \eta) + S_2 J_1(u_2 \eta)}{S_1 K_0(u_2 \eta) + S_2 J_0(u_2 \eta)} = \frac{K_1(u_3)}{K_0(u_3)}. \quad (16)$$

We can also define a new $V$ parameter ($\tilde{V}_{23}$) to characterize the index step at the cladding-jacket interface.

$$\tilde{V}_{23}^2 = \left( \frac{2\pi a}{\lambda} \right)^2 (n_2^2 - n_3^2) = \left( \frac{u_3}{\eta} \right)^2 - u_2^2. \quad (17)$$

For infinite-cladding gain-guided (IC-GG) fibers in subsection 2.A, all the information required to determine $u_1$ and $u_2$, and therefore the mode field profile was encoded in a single dimensionless complex parameter $\tilde{V}_{12}$. In the case of FC-GG fibers, in addition to $\tilde{V}_{12}$, we also need the values of the dimensionless parameters $\tilde{V}_{23}$ and $\eta$, in order to determine $u_1$, $u_2$, and $u_3$, required for the mode profile. Eqs. 8, 16, and 17 will be used to solve for the modal parameters, given the values of $\tilde{V}_{12}$, $\tilde{V}_{23}$, and $\eta$.

## 3. Mode profile in a FC-GG fiber

The formulation we presented in section 2 is applicable in general to any selection of complex values for the refractive indexes of the core, cladding, and jacket. However, we focus our analysis on the case where the cladding and jacket are not active, so $n_2$ and $n_3$ are assumed to be real throughout this paper. This practical choice conforms with the assumptions of Ref. [3] and also the subsequent experimental studies of GG+IAG optical fibers. We also note that for simplicity, we choose the jacket to be made of air ($n_3 = 1$) to carry out our analysis. However, we will comment on other choices for the jacket material.

The electric field profile of the LP01 mode for the case of an IC-GG fiber was analyzed in detail by Siegman [3]. In Fig. 2, we present the amplitude of the electric field profiles for a generic example where $\Delta N = -20$, for the case of infinite cladding ($\eta = \infty$), as well as for three cases of finite cladding diameters, characterized by $\eta = 3, 2, 1.5$. In each case, the amplitude of the LP01 mode is plotted for three values of gain $G = 2G_{th}$ (solid), $G = G_{th}$ (dashed), and $G = 0$ (dot-dashed). Point A, marked on Fig. 1, represents $\Delta N = -20$ and $G = G_{th}$, for the IC case. We note that we use an identical value of $G_{th}$ at $\Delta N = -20$ as evaluated for IC-GG fibers for all four plots, since the concept of a guidance threshold might be meaningless in FC-GG fibers, as will be argued later in this section.

We first recall the behavior of the field profile for infinite cladding ($\eta = \infty$). The point characterized by $G = 2G_{th}$ belongs to the parameter region $G > G_{th}$, generally associated with an intensity tail in the cladding, which drops exponentially in the radial direction. As the gain parameter ($G$) approaches the threshold value ($G_{th}$), the tail extends farther into the cladding and the rate of exponential decay slows down. At $G = G_{th}$, the tail reaches a finite non-zero asymptotic value at infinity. For $G < G_{th}$, including $G = 0$, the cladding tail grows exponentially into the cladding region and the field turns into a leaky mode. This behavior is clearly observed in the top-left subfigure of Fig. 2.

We now investigate the LP01 field profiles for the cases of finite cladding: $\eta = 3$, $\eta = 2$, and $\eta = 1.5$, presented in the top-right, bottom-left, and bottom-right subfigures of Fig. 2, respectively. We expect that the electric field profile of an LP01 mode in a FC-GG fiber to somewhat resemble that of an IC-GG fiber in the core and the cladding. The profiles in Fig. 2 show that this expectation holds true and is most accurate for large values of $\eta$, such as $\eta = 3$. The most notable departure of the filed profiles of the FC-GG fibers from that of IC-GG is the strong truncation of the intensity tail at the cladding-jacket interface, due to the large index-step. For $G \leq G_{th}$, the truncation has a greater impact on the profile of the mode in the core and in the cladding, compared with $G > G_{th}$. The impact is also greater for smaller values of $\eta$, as expected. We note that except for the case of $\eta = \infty$, all the mode profiles plotted in Fig. 2 satisfy the normalization condition of Eq. 5, if the vertical axes in Fig. 2 are to be regarded in unit of $(1/a)$. However, since the two cases of $G = 0$ and $G = G_{th}$ are not normalizable for $\eta = \infty$ in the top-left figure, we scaled all the profiles belonging to $\eta = \infty$ to match that of $\eta = 3$ at the origin ($r = 0$), for easier comparison.

Another important point for FC-GG profiles is that the truncation at the cladding-jacket interface turns leaky modes into proper normalizable guided modes. In other words, unlike the IC-GG fiber, $\Re(u_2) = 0$ no longer serves as a GG threshold for FC-GG fibers. We note that the normalizability requirement that imposed a condition of $\Re(u_2) > 0$ on the guided modes of IC-GG fibers, is now replaced by $\Re(u_3) > 0$ for FC-GG fibers. Those solutions of Eqs. 8, 17, and 16 that satisfy $\Re(u_3) < 0$ are the leaky modes which grow exponentially to infinity in the transverse direction in the jacket. In the next paragraph, we will argue that for a sufficiently large refractive index contrast between the cladding layer and the jacket, it is possible for all FC-GG modes defined on the complex plane of $(\Delta N, G)$ to be guiding. In other words, the condition $\Re(u_3) = 0$ may no longer translate into a guiding threshold on this complex plane and there will be no leaky LP01-like modes. However, for finite values of $\eta$, it might be interesting to find the quasi-threshold line of $\Re(u_2) = 0$. This quasi-threshold lines are plotted on Fig. 1 for $\eta = 2$ and $\eta = 1.5$ as dashed and dot-dashed, respectively. A quasi-threshold line for $\eta > 3$ becomes virtually indistinguishable from the solid threshold line ($\eta = \infty$) belonging to IC-GG fibers in this figure. We emphasize again that the quasi-threshold lines are for illustration purposes and do not present any physical threshold for guiding versus leaky properties in FC-GG fibers.

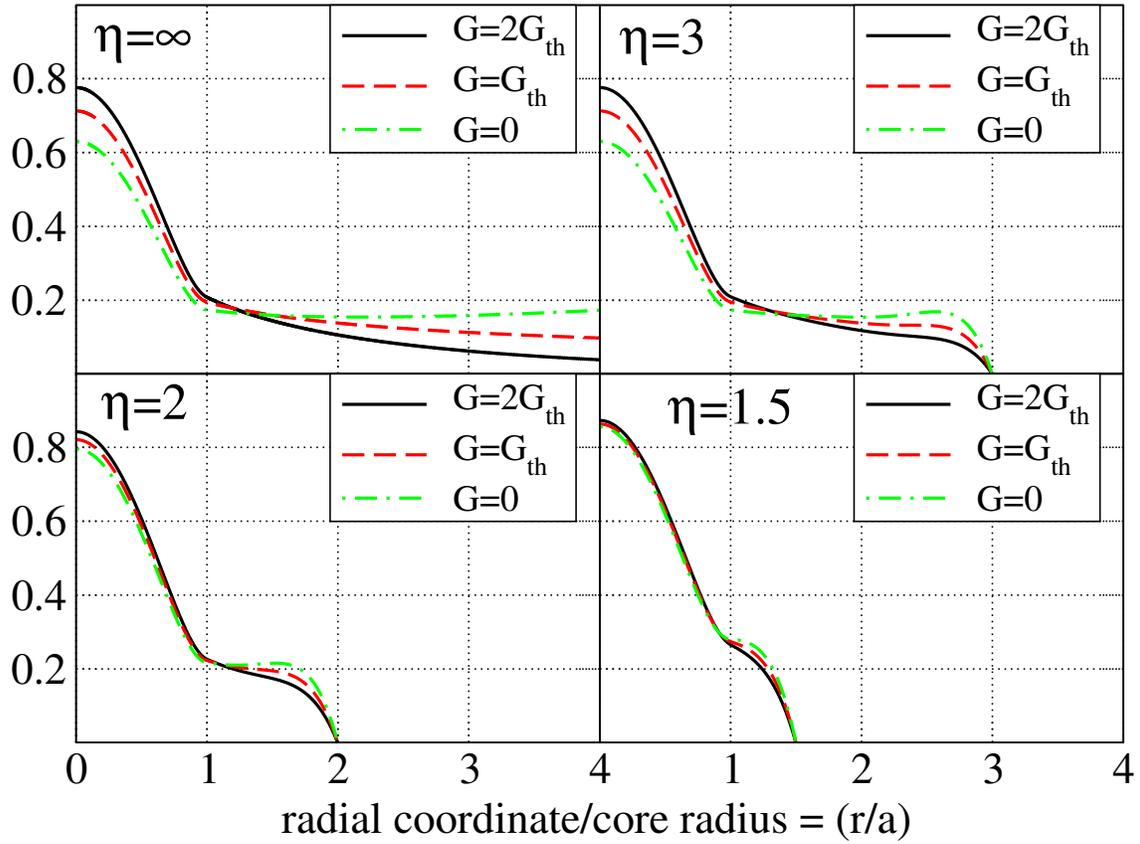

Fig. 2. Electric field profiles for the case of IC-GG fibers ($\eta = \infty$, top-left), and FC-GG fibers ($\eta = 3,\ 2,\ 1.5$) in the other 3 figures.

We now argue that $\Re(u_3) = 0$ may no longer translate into a guiding threshold. In practice, there is an often large refractive index contrast between the cladding layer and the jacket, especially if the jacket is assumed to be air (our choice in this paper). Therefore, $\tilde{V}_{23}$ can be quite large. Using Eq. 17 and assuming $\tilde{V}_{23} \gg |u_2|$, we get

$$\Re(u_3) \approx \eta \tilde{V}_{23} \left( 1 - \frac{\Im(u_2)^2 - \Re(u_2)^2}{2\tilde{V}_{23}^2} + O(\tilde{V}_{23}^{-4}) \right). \tag{18}$$

It is clear that unless $\Im(u_2)^2 - \Re(u_2)^2 \sim 2\tilde{V}_{23}^2$, $\Re(u_3)$ remains positive and the mode is confined. Therefore, unlike the IC-GG fibers, guided modes exist even below the solid-line threshold of Fig. 1.

We note that for all three choices of $\eta$ studied in this paper, the values of the modal parameters $u_1$ and $u_2$ are almost the same in IC-GG and FC-GG fibers. Therefore, $u_1$ and $u_2$ as the numerical solutions of the much simpler IC-GG dispersion equation 9 can be used as a starting point in the search algorithm to find $u_1$, $u_2$, and $u_3$ in the case of FC-GG fibers. For our choice of parameters, $\eta = 3$ is practically identical to $\eta = \infty$, when it comes to solving for $u_1$ and $u_2$ for the root-finding algorithm we employed in Mathematica. We note that solving for the modal parameters, using the dispersion equation of the FC-GG becomes increasingly harder as $\eta$ gets larger.

Lastly, we would like to comment on our choice of $n_3 = 1$. Our observations reported in this paper should hold true as long as $n_2$ and $n_3$ are not too close. On the other hand, in the limit of $n_3 = n_2$, the FC-GG fiber becomes identical to the IC-GG case. In terms of using a root-finding algorithm with $u_1$ and $u_2$ from IC-GG as the seed for the FC-GG case, our choice of $n_3 = 1$ is the most stringent choice and larger values of $n_3$ will result in even greater resemblance between the LP01 profiles and modal parameters in IC-GG and FC-GG fibers.

## 4. Beam quality factor in a FC-GG fiber

The reported values of the beam quality factor $M^2$ in several experiments [4–8] are notably larger than unity, in the range of 1.05-2.0, even in fibers that are designed to operate as single-mode. This should not be surprising considering the substantial departure of the LP01 profile of a GG+IAG fiber from a Gaussian-like beam, as similarly reported in other unconventional optical fibers [10]. We recently conducted an extensive theoretical study of the beam quality factor $M^2$ of single-mode fibers in the presence of gain [9]. We showed that the theoretical predictions of the $M^2$ in the GG+IAG region can be substantially larger than the experimental measurements ($M^2 > 10$), and attributed this difference to the truncation of the long tail of the beam extending all the way into the cladding region in the experiments, which can lower $M^2$ substantially. For example, Ref. [4] reports $M^2 \leq 2$ for a GG+IAG optical fiber with a 100 $\mu m$ core diameter and a 250 $\mu m$ cladding diameter. Similarly, Ref. [5] reports $1.2 \leq M^2 \leq 1.5$, where the core and cladding diameters are 200 $\mu m$ and 340 $\mu m$, respectively. Ref. [8] reports $M^2 \approx 1.4$ for a GG+IAG optical fiber with a 100 $\mu m$ core diameter and a 250 $\mu m$ cladding diameter. The cladding diameter is not much larger than the core diameter in either experiments. Therefore, it is not surprising that these experiments measure such low values of $M^2$.

In order to explore the impact of the long intensity tail in the cladding on the value of $M^2$ in Ref. [9], we introduced a Gaussian apodization to softly truncate the long intensity tail of the beam calculated from an IC-GG fiber. Although this served as a reasonable and convenient approximation, the mode profiles plotted in Fig. 2 show that a more rigorous study is warranted. For example, a simple Gaussian field truncation at the cladding-jacket boundary for $\eta = 1.5$ does not seem be an adequate approximation.

Our method for calculating the $M^2$ is similar to that reported in Ref. [9]. Consider an optical beam with the electric field profile $E(x, y, z)$ propagating in the $z$ direction. The beam center $\langle x \rangle$ and the standard deviation of the intensity distribution $\sigma_x^2$ across the $x$ coordinate are

$$\langle x \rangle(z) = \iint dS \ x \ |E(x,y,z)|^2, \tag{19}$$

$$\sigma_x^2(z) = \iint dS \ \left(x - \langle x \rangle(z)\right)^2 \ |E(x,y,z)|^2. \tag{20}$$

Since we only consider cylindrically symmetric optical fibers, the results are identical for the $y$ coordinate, and we take the liberty in dropping the $x$ subscript (e.g. $M^2$) when convenient. It can be shown that the standard deviation in Eq. 20, in the paraxial approximation, obeys a universal free-space propagation rule of the form

$$\sigma_x^2(z) = \sigma_x^2(z_0) + A\frac{\lambda}{2\pi}(z - z_0) + B\frac{\lambda^2}{4\pi^2}(z - z_0)^2. \tag{21}$$

$z_0$ is the coordinate of the output facet of the fiber, which does not necessarily coincide with the position of the beam waist. Ref. [11] has shown that

$$A = -i \iint dS\big(x - \langle x \rangle(z_0)\big)\Big[E\frac{\partial E^\star}{\partial x} - c.c.\Big], \tag{22}$$

$$B = \iint dS \left|\frac{\partial E}{\partial x}\right|^2 + \frac{1}{4}\Big[\iint dS(E\frac{\partial E^\star}{\partial x} - c.c.)\Big]^2, \tag{23}$$

where $E \equiv E(x, y, z_0)$ is implied in Eqs. 22 and 23. The position of the beam waist $\tilde{z}_{0x}$ and the beam-quality factor $M_x^2$ are given by

$$\tilde{z}_{0x} = z_0 - \pi A/(\lambda B), \tag{24}$$

$$M_x^2 = \sqrt{4B\sigma_x^2(z_0) - A^2}. \tag{25}$$

We note that we slightly differ with Ref. [11] in the sign of the frequency term and also the definition of the $A$ term. Our definitions remain consistent with Ref. [9], except for a missing minus sign in the definition of $A$, which has been corrected here in Eq. 22. For the case of finite cladding, we consider the generic electric field profile of Eq. 11. While the value of $M^2$ in Eq. ?? can be evaluated numerically, we also derive a closed-form analytical expression for the $M^2$ parameter in the Appendix.

The contour plots of $M^2$ as a function of $\Delta N$ and $G$ are shown in Figs. 3(a), 3(b), 3(c), 3(d). Fig. 3(a) relates to $\eta = \infty$ representing the IC-GG-IAG fibers. The dashed line in this figure identifies the threshold values of the dimensionless gain parameter $G$ required to produce a confined and amplifying LP01 mode as also shown in the form of a solid line in Fig. 1. Figs. 3(b), 3(c), 3(d) relate to $\eta = 3$, $\eta = 2$, and $\eta = 1.5$, respectively, representing different FC-GG-IAG fibers.

The results in Fig. 3(a), related to the GG+IAG region of FC-GG fibers, show that $M^2$ is quite large over the entire parameter space and becomes exceptionally large near the (dashed) LP01 threshold. The large value of $M^2$ especially near the threshold is the result of the long tail of the beam intensity extending all the way into the cladding region [9]. The situation is quite different for FC-GG fibers as shown in Figs. 3(b), 3(c), 3(d). The truncation of the beam tail at the cladding-jacket interface results in a substantial reduction in the calculated value of $M^2$. The above observations are quite important in relating the $M^2$ values reported in this paper to the experimental measurements.

We note that in the GG-IAG region ($\Delta N < 0$), the beam quality factor $M^2$ increases as $|\Delta N|$ becomes progressively larger. This trend can be observed in Figs. 3(b), 3(c), 3(d). Also, in each of these figures, the contour lines evenually become vertical for large enough values of $|\Delta N|$. This is almost starting to happen for $\eta = 1.5$ in Fig. 3(d) at $\Delta N \approx -100$. For larger values of $\eta$, this effect happens at larger values of $|\Delta N|$. In other words, for very negative values of $\Delta N$, the $M^2$ parameter becomes independent of the $G$ parameter. This can be seen in Fig. 4, where we plot $M^2$ as a function of $G$ for $\Delta N = -100$ (left) and $\Delta N = -1000$ (right). It can also be seen that for each of the three choices of $\eta = 1.5$ (solid), $\eta = 1.7$ (dashed), and $\eta = 1.8$ (dot-dashed), a larger value of $\Delta N$ results in a larger beam quality factor. This is also consistent with our remarks regarding the extrapolation of Figs. 3(b), 3(c), 3(d) to more negative values of $\Delta N$. It can also be seen (as also shown in Figs. 3(b), 3(c), 3(d)) that $M^2$ increases with the value of $\eta$. As we stated before, solving for the modal parameters $u_1$, $u_2$, and $u_3$ becomes exceedingly difficult for large values of $\eta$, even if we start with the modal parameters $u_1$ and $u_2$ from the much simpler IC-GG dispersion equation 9. For $\eta > 2$ and $\Delta N \ll -100$, finding a solution to Eq. 16 becomes a complicated numerical problem and requires accuracy beyond the machine precision. Addressing such numerical issues are beyond the scope of this paper and we have chosen to extract the expected values of $M^2$ for $\eta > 2$ and $\Delta N \ll -100$ by observing the trends in our figures.

## 5. Conclusions

In a previous paper [9], we showed that $M^2$ can be substantially larger than unity even for single-mode fibers in the GG+IAG region. We argued that the truncation of the beam profile due to a finite cladding diameter is responsible for the lower $M^2$ values observed in several experimental measurements. Therefore, in Ref. [9], we introduced a Gaussian apodization to softly truncate the long intensity tail of the beam and explore its impact on $M^2$ values. Here, for the first time, we present a rigorous analysis of the beam profile and $M^2$ for GG-IAG fibers with finite cladding diameters. Our analytical derivations presented in the Appendix are generally applicable to any step-index optical fiber with a finite cladding diameter and gain in the core, such as the conventional double-cladding fiber lasers.

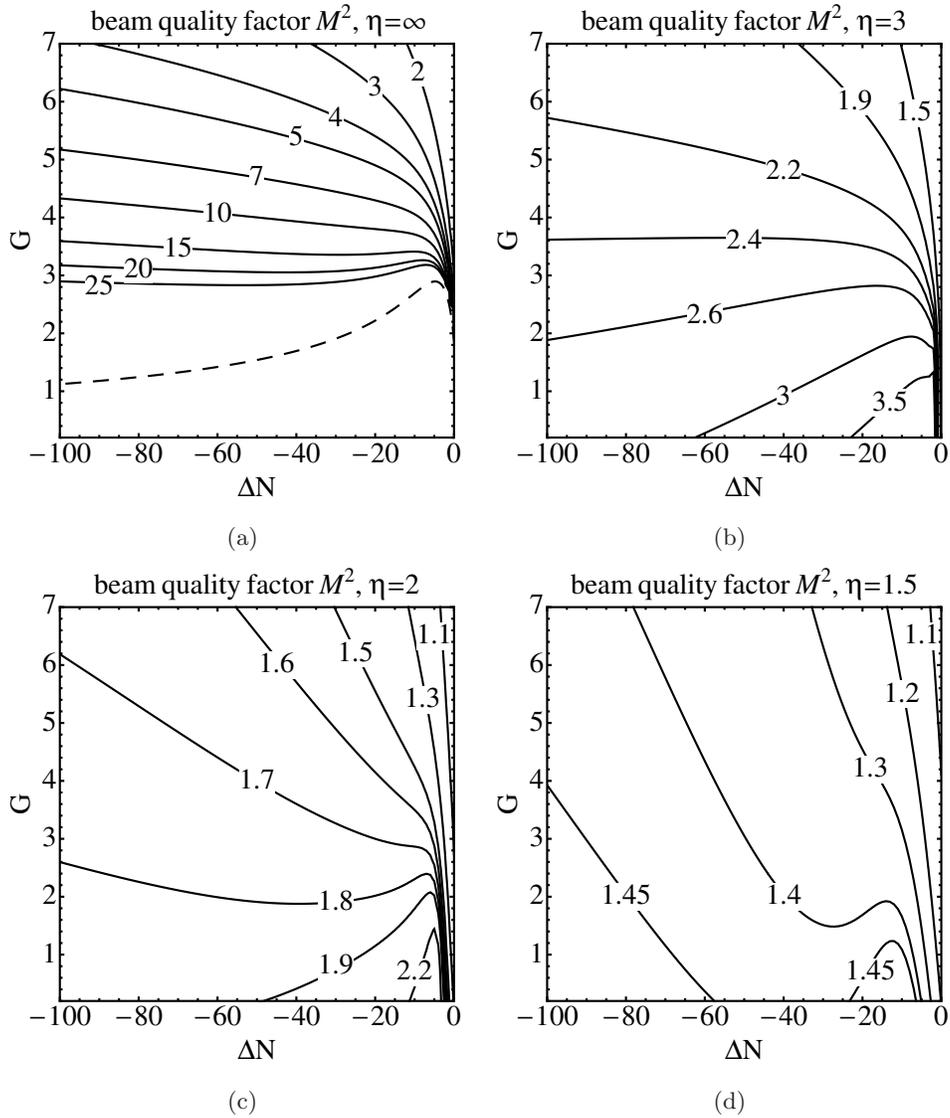

Fig. 3. Contour plot of $M^2$ as a function of $\Delta N$ and $G$ in the GG+IAG region for a) IC-GG fibers with $\eta = \infty$, b), c) , and d) for FC-GG fibers with $\eta = 3$, $\eta = 2$, and $\eta = 1.5$, respectively.

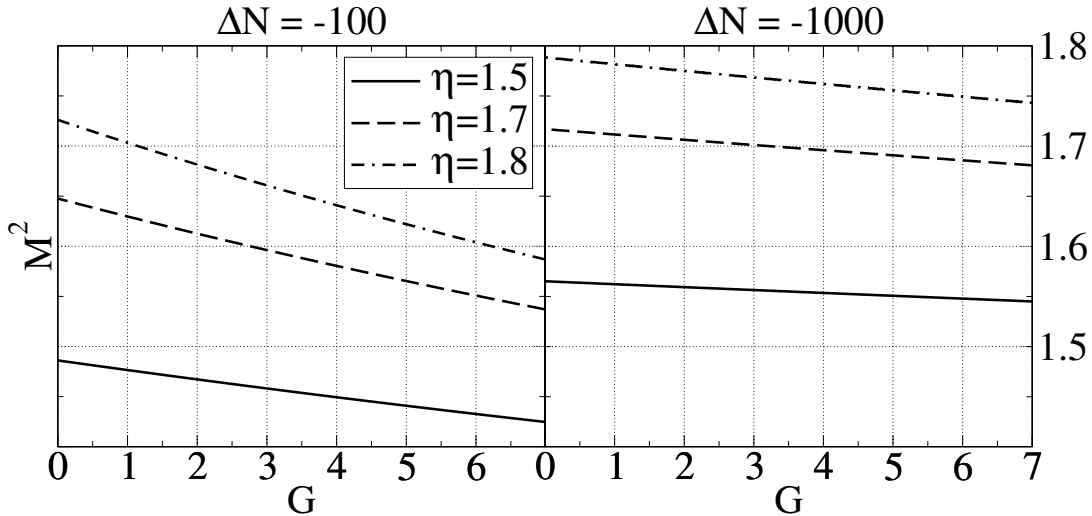

Fig. 4. $M^2$ as a function of $G$ for $\Delta N = -100$ (left) and $\Delta N = -1000$ (right). The lines in each figure relate to the different values of $\eta = 1.5$ (solid), $\eta = 1.7$ (dashed), and $\eta = 1.8$ (dot-dashed).

The reported values of $M^2$ in several experiments on GG+IAG fiber lasers are in the range of 1.05-2.0 [4–8]. For example, Ref. [4] reports $M^2 \leq 2$ for a GG+IAG optical fiber with a 100 $\mu m$ core diameter and a 250 $\mu m$ cladding diameter ($\eta = 2.5$). Similarly, Ref. [5] reports $1.2 \leq M^2 \leq 1.5$, where the core and cladding diameters are 200 $\mu m$ and 340 $\mu m$, respectively, where $\eta = 1.7$. These measurements are quantitatively consistent with our results presented in Fig. 4, for $\Delta N \leq -1000$. Ref. [8] reports $M^2 \approx 1.4$ for a GG+IAG optical fiber with a 100 $\mu m$ core diameter and a 250 $\mu m$ cladding diameter ($\eta = 2.5$). This value of $M^2$ is slightly lower than what we expect from our simulations. However, there are uncertainties associated with $M^2$ measurements, as is clear from the difference between the reported results of Ref. [4] and Ref. [8]. Therefore, it is reasonable to expect some slight differences between theory and experiment.

GG-IAG fibers provide a rich parameter space to explore and design unconventional optical-fiber-based devices. Unfortunately, the pump confinement is a serious issue in such fibers, since the pump light is expelled from the core, because of its index-antiguiding behavior. A practical solution to the pump confinement problem will pave the way to interesting device-level applications of such fibers.

## A. Appendix

We present closed-form analytical expressions for the parameters defined in Eqs. 22 and 23, which can be used to evaluate the $M^2$ parameter. We first introduce the following functions and functionals:

$$\zeta_{m,n;\tilde{u}}^{p,q} = (p\tilde{u}^2 + q\tilde{u}^{\star 2})^m / (p\tilde{u}^2 - q\tilde{u}^{\star 2})^n, \tag{26}$$

$$F_n(x, X, Y) = x^\star X_n(x) Y_n'(x^\star) - x X_n'(x) Y_n(x^\star) \tag{27}$$

$$G_n(x, X, Y) = X_n(x) Y_n(x^\star) \tag{28}$$

$$P_n(M, \tilde{u}, \tilde{a}, X, Y, p, q) = M\tilde{a}^2\ \zeta_{0,1;\tilde{u}}^{p,q}\ F_n(\tilde{u}, X, Y). \tag{29}$$

$$Q_n(M, \tilde{u}, \tilde{a}, X, Y, p, q) = \tag{30}$$
$$2M\tilde{a}^4\ \zeta_{1,2;\tilde{u}}^{p,q}\ G_n(\tilde{u}, X, Y)$$
$$+ 4M|\tilde{u}|^2\tilde{a}^4\ \zeta_{0,2;\tilde{u}}^{p,q}\ G_{n+1}(\tilde{u}, X, Y)$$
$$+ \tilde{a}^2\left(1 - 4\zeta_{1,2;\tilde{u}}^{p,q}\right)\ P_n(M, \tilde{u}, \tilde{a}, X, Y, p, q).$$

$$R_n(M, \tilde{u}, \tilde{a}, X, Y, p, q) = M\tilde{a}^2\ F_n(\tilde{u}, X, Y) \tag{31}$$
$$- \tilde{a}^{-2}\zeta_{1,0;\tilde{u}}^{p,-q}\ Q_n(M, \tilde{u}, \tilde{a}, X, Y, p, q)$$

$$U_n(M, \tilde{u}, \tilde{a}, X, Y, p, q) = \tilde{a}^{-2}|\tilde{u}|^2 P_n(M, \tilde{k}, \tilde{a}, X, Y, p, q) \tag{32}$$

$$S_n(I, M, \tilde{u}, X, Y, p, q) = I_n(M, \tilde{u}, b, X, Y, p, q) \tag{33}$$
$$- I_n(M, \tilde{u}, a, X, Y, p, q).$$

$$T_n(I) = \tag{34}$$
$$I_n(|\tilde{A}|^2, u_1, a, J, J, 1, 1) - I_n(|\tilde{D}|^2, u_3, b, K, K, -1, -1) +$$
$$S_n(I, |\tilde{B}|^2, u_2, J, J, 1, 1) + S_n(I, \tilde{B}\tilde{C}^\star, u_2, J, K, 1, -1) +$$
$$S_n(I, \tilde{C}\tilde{B}^\star, u_2, K, J, -, +) + S_n(I, |\tilde{C}|^2, u_2, K, K, -, -).$$

In the above expressions, $p$ and $q$ each can take the values of $\pm 1$, which are shortened to $\pm$ in Eq. 34. $X_n$ and $Y_n$ each represent one of the Bessel functions $J_n$ and $K_n$, respectively. We also have:

$$\tilde{A} = \frac{1}{J_0(u_1)}, \quad \tilde{B} = \frac{1}{L_0(u_2)}\frac{S_2}{S_1}, \quad \tilde{C} = \frac{1}{L_0(u_2)}, \tag{35}$$
$$\tilde{D} = \frac{L_0(u_2\eta)}{L_0(u_2)}\frac{1}{K_0(u_3)}.$$

Equipped with these functions and functionals, we can obtain the following explicit expressions for the parameters of interest

$$\frac{1}{\tilde{N}^2} = 2\pi T_0(P), \tag{36}$$

$$\sigma_x^2(z_0) = \pi \tilde{N}^2 T_0(Q), \tag{37}$$

$$A(z_0) = i\pi \tilde{N}^2 T_0(R/2), \tag{38}$$

$$B(z_0) = \pi \tilde{N}^2 T_1(U). \tag{39}$$

We note that the analytical derivations involve the evaluation of complex-valued moments of products of Bessel functions, which are beyond the scope of the present paper. To the best knowledge of the authors, the techniques employed to derive these analytical expressions are novel and will be presented in a future publication.

## Acknowledgments

The authors achnowledge support from the UWM Research Growth Initiative grant for this publication.

## References

1. A. Mafi, J. V. Moloney, D. Kouznetsov, A. Schülzgen, S. Jiang, T. Luo, and N. Peyghambarian, "A large-core compact high-power single-mode photonic crystal fiber laser," IEEE Photon. Tech. Lett. **16**, 2595–2597 (2004).
2. A. E. Siegman, "Propagating modes in gain-guided optical fibers," J. Opt. Soc. Am. A **20**, 1617–1628 (2003).
3. A. E. Siegman, "Gain-guided, index-antiguided fiber lasers," J. Opt. Soc. Am. B **24**, 1677–1682 (2007).
4. A. E. Siegman, Y. Chen, V. Sudesh, M. Richardson, M. Bass, P. Foy, W. Hawkins, and J. Ballato, "Confined propagation and near single mode laser oscillation in a gain guided, index antiguided optical fiber," Appl. Phys. Lett. **89,** 251101–251103 (2006).
5. V. Sudesh, T. McComb, Y. Chen, M. Bass, M. Richardson, J. Ballato, and A. E. Siegman, "Diode-pumped 200 $\mu$m diameter core, gain-guided, index-antiguided single mode fiber laser," Appl. Phys. B **90,** 369–372 (2008).
6. Y. Chen, V. Sudesh, T. McComb, M. Richardson, M. Bass, and J. Ballato, "Lasing in a gain-guiding index antiguided fiber," J. Opt. Soc. Am. B **24,** 1683–1688 (2007).
7. Y. Chen, T. McComb, V. Sudesh, M. Richardson, and M. Bass, "Very large-core, single-mode, gain-guiding, index-antiguided fiber lasers," Opt. Lett. **32,** 2505–2507 (2007).
8. W. Hageman, Y. Chen, X. Wang, L. Gao, G. U. Kim, M. Richardson, and M. Bass, "Scalable side-pumped, gain-guided index-anti-guided fiber laser," J. Opt. Soc. Am. A **27**, 2451–2459 (2010).
9. K. M. Gundu, P. Gandomkar Yarandi, and A. Mafi, "Beam quality factor of single-mode gain-guiding fiber lasers," Opt. Lett. **35,** 4124–4126 (2010).

10. A. Mafi and J. V. Moloney, "Beam Quality of Photonic-Crystal Fibers," J. Lightwave Technol. **23,** 2267–2270 (2005).
11. H. Yoda, P. Polynkin, and M. Mansuripur, "Beam quality factor of higher order modes in a step-index fiber," J. Lightwave Technol. **24**, 1350–1355 (2006).